\documentclass[12pt]{article}
\usepackage{epsf}
\usepackage{hyperref}
\usepackage{graphicx, rotating}
\usepackage{subfigure}
\usepackage[margin=2.5cm]{geometry}
\usepackage{amsmath}
\usepackage{footnote}
\usepackage[bottom]{footmisc}
\usepackage{threeparttable}
\usepackage{caption}
\usepackage{amsthm}
\usepackage{enumitem}
\usepackage{graphics}
\usepackage{float}
\usepackage{amssymb}
\usepackage{natbib}
\usepackage{lscape}
\usepackage{array}
\usepackage{setspace}
\usepackage{verbatim}
\usepackage{multirow}

\bibpunct{(}{)}{;}{a}{,}{,}

\begin{document}
\title{\LARGE Survival Trees for Interval-Censored Survival data}
\author{Wei Fu, Jeffrey S. Simonoff\\
              New York University}
\date{\today}
\maketitle

\begin{abstract}
Interval-censored data, in which  the event time is only known to lie in some time interval, arise commonly in practice; for example, in a medical study in which patients visit clinics or hospitals at pre-scheduled times, and the events of interest occur between visits. Such data are appropriately analyzed using methods that account for this uncertainty in event time measurement. In this paper we propose a survival tree method for interval-censored data based on the conditional inference framework. Using Monte Carlo simulations we find that the tree is effective in uncovering underlying tree structure, performs similarly to an interval-censored Cox proportional hazards model fit when the true relationship is linear, and performs at least as well as (and in the presence of right-censoring outperforms) the Cox model when the true relationship is not linear. Further, the interval-censored tree outperforms survival trees based on imputing the event time as an endpoint or the midpoint of the censoring interval. We illustrate the application of the method on tooth emergence data.
\end{abstract}


\section{Introduction}
In classic time-to-event or survival data analysis, the object of interest is the occurrence time of the event, which is usually observed or right-censored. Such right-censored data are well-studied and there are numerous methods, including (semi-)parametric and nonparametric methods, available to handle such data. However, there are many other incomplete data scenarios in survival analysis, one being interval-censored data \citep{ICbook}.

Interval-censored (IC) data arise commonly in a medical or longitudinal study in which the subjects are assessed periodically. For example, patients often visit clinics or hospitals at pre-scheduled times, and the events of interest may occur between visits. In this situation, the event time is only known to lie in some time interval. Such data are called interval-censored data, while the occurrence times of the event are said to be interval-censored. Note that right-censoring is a special case of interval-censoring.

Because of the relative lack of well-established techniques for dealing with interval-censored data, an \textit{ad hoc}  approach is to assume that the event occurred at the middle (or end) of the time interval. However, such an approach is known to bias the results and lead to invalid inferences \citep{lindsey1998methods}.

In this paper, we propose a nonparametric recursive-partitioning (tree) method appropriate for interval-censored data. The goal of this tree algorithm is to form groups of subjects within which subjects have similar survival distributions, thereby segmenting the population. The proposed method is an extension of the survival tree method proposed by \cite{Hothorn06} (which is designed to handle right-censored survival data), which was adapted to left-truncated and right-censored data in \cite{Fu2017}.

\section{An interval-censored survival tree}

\cite{Hothorn06} presented a framework embedding recursive partitioning into a well-defined theory of permutation tests developed by \cite{strasser99onthe}. In the popular tree algorithm CART \citep{cart84}, the selection of the splitting variable and the selection of the splitting point are accomplished in one step. Such a procedure results in the method being more likely to split on attributes with more possible split points, and hence is biased in terms of selection of the splitting variable.

The conditional inference tree algorithm of \cite{Hothorn06}
addresses this problem by separating these two steps. The algorithm
works by first selecting the splitting variable, through the use of
a conditional distribution that is constructed based on the
assumption that the response and the covariates are independent.
After the splitting variable is selected, the split point can be
determined by any criterion, including those discussed by
\cite{cart84}. The conditional inference tree algorithm that
implements this method is implemented in the \texttt{ctree} function
in the R package \texttt{partykit} \citep{partykit}.

\subsection{Extending the survival tree of \cite{Hothorn06}}

As far as we are aware, the only other proposal of a survival tree method for interval-censored data was made in \cite{YinAnderson02}. This is based on a formulation consistent with that of CART \citep{cart84}, and would presumably therefore suffer from splitting variable bias as noted above. There does not appear to be any publicly-available software to implement this proposal.

The conditional inference tree of \cite{Hothorn06} measures the
association of $Y$ and a predictor $X_j$ based on a random sample $\mathcal{L}_n$ by linear statistics of the
form
\[
T_j(\mathcal{L}_n,w) = vec \left(\sum_{i=1}^n w_ig_i(X_{ji})h \left(
Y_i\right)^T\right)\in \mathbb{R}^{p_jq}                       \]
where $g_j : \mathcal{X}_j \to \mathbb{R}^{p_j} $ is a nonrandom
transformation of covariate $X_j$ and $h : \mathcal{Y} \times
\mathcal{Y}^n \to \mathbb{R}^q$ is the influence function of the
response $Y$. In the case of interval-censoring, the response is $Y_i = (L_i,
R_i]$, where $(L_i, R_i]$ is the censoring interval within which the
event time lies, and $h \left( Y_i\right) = U_i$, the log-rank
score. The function of the log-rank score in the algorithm is to
assign the univariate scalar value $U_i$ to the bivariate response
$Y_i = (L_i, R_i]$, so the algorithm can then execute in the same
way as in the univariate numeric response case. Each $T_j(\mathcal{L}_n,w)$ is
standardized using the conditional expectation $\mu_j$ and
covariance $\Sigma_j$ of $T_j(\mathcal{L}_n,w)$ given by
\cite{strasser99onthe} (fixing the covariates and conditioning on all possible permutations of the responses), and the algorithm picks the covariate $X_j$
associated with the smallest $p$-value as the splitting variable, stopping splitting when all $p$-values are above a threshold (.05 by default).

The log-rank score was first proposed by \cite{peto1972asymptotically}, who derived general (asymptotically efficient) rank invariant test procedures for detecting differences between two groups of independent observations.
They also established that under $H_0$, the null hypothesis that groups have the same distribution, using the log-rank score statistics and using the difference between the observed and expected event counts at event times (which are used by the log-rank test) to describe a group are equivalent, which means that using the log-rank score and using the log-rank test to compare survival curves of different groups are equivalent, under the condition of independent observations. More details about the log-rank score and its application in survival trees can be found in \cite{Fu2017}.

The log-rank score for interval-censored data can be easily derived from the score equation given in \cite{pan1998rank}, who extended the rank invariant tests of \cite{peto1972asymptotically} to left-truncated and interval-censored data. The log-rank score for interval-censored data is given by
\begin{equation}\label{equ1}
U_i = \frac{\hat{S}(L_i) \log \hat{S}(L_i) - \hat{S}(R_i) \log \hat{S}(R_i)}{\hat{S}(L_i)-\hat{S}(R_i)},
\end{equation}
where $L_i$ and $R_i$ are the lower and upper boundaries of the censoring interval for the $i$th observation, respectively. Note that $\hat{S}$ is the nonparametric maximum likelihood estimator (NPMLE) of the survival function. In practice, such an estimator can be constructed using the algorithm as proposed by \cite{turnbull1976empirical}. The estimator uses a self-consistency argument to motivate an iterative algorithm for the NPMLE, which turns out to be a special case of the EM-algorithm. The estimator simplifies to the Kaplan-Meier estimator when event and censored times are known exactly, and is implemented in the \texttt{icfit} function in the R package \texttt{interval} \citep{Rinterval}.

A special case is when the event time is observed, so interval $(L_i, R_i]$ reduces to a point since $L_i = R_i$.
In this case, equation (\ref{equ1}) cannot be used directly to compute the log-rank score since $\hat{S}(L_i)= \hat{S}(R_i)$. Note that in such a case equation (\ref{equ1}) can be seen as the derivative of the function $S\log S$ at $S=\hat{S}(L_i)$, and therefore the corresponding log-rank score is
\[ U_i = 1 + \log \hat S(L_i) \hspace{0.1in} \text{if} \hspace{0.1in} \delta_i=1\]
and
\[ U_i = \log \hat S(L_i) \hspace{0.1in} \text{if} \hspace{0.1in} \delta_i=0.\]

\section{Properties of the tree method}
In this section, we use computer simulations to investigate the
properties of the proposed tree method. We assume that the event
time $T$ is generated from distribution $F(t)$. To generate the
censoring interval under the non-informative censoring assumption,
we generate the censoring mechanism of $T$ to mimic a longitudinal
study. Suppose there are $k+1$ examination times $\{0, t_1,
t_2,...,t_k \}$, which segment the time line into $k+1$ time
intervals $(0,t_1], (t_1,t_2],...,(t_k,\infty)$. The censoring
interval of $T$ is the one that contains $T$. Note that $T$ is
generated independently from the censoring mechanism. The gap
between two examination times $\delta_t = t_{j}-t_{j-1}$ can be
fixed or be a random variable from some distribution $G(t)$. In
either case, this mechanism ensures the possibility that some
observations can potentially be right-censored, i.e. $T$ lies in
$(t_k,\infty)$. This mechanism is used in \cite{pan2000multiple}.

We will study the properties of the proposed tree method in terms of its unbiasedness in selecting the splitting variable, its ability to recover the correct tree structure, and its prediction
performance. The simulation setups are similar to those in \cite{Fu2017}.

\subsection{Unbiasedness of variable selection}
The survival tree of \cite{Hothorn06} is unbiased in terms of selecting the splitting variable, which
means that it selects each covariate with equal probability of splitting under the condition of
independence between response and covariates. This suggests that the proposed interval-censored (IC) tree
based on it is also unbiased. We explore the unbiasedness of the proposed IC tree in this section.

The event time $T$ is randomly generated with the following possible distributions:
\begin{itemize}
    \item Exponential with rate $1/3.2$
    \item Weibull distribution with shape $=0.8$, scale $=3$
    \item Lognormal distribution with mean $=0.8$, standard deviation $=1$
\end{itemize}

The censoring interval is generated as described above for each generated $T$, where $k = 5$ and $\delta_t$ is either fixed or a random variable. Note that the value of $\delta_t$ ($\delta_t$ is fixed) or the distribution of $\delta_t$ ($\delta_t$ is a random variable) controls the proportion of observations being right-censored. For each case, we select the parameters such that $20\%$ and $40\%$ of the observations are right-censored in the light and heavy censoring cases, respectively.

The observed response for each observation is the censoring interval $(L,R]$. There are five independent covariates $\{X_1, X_2, X_3, X_4, X_5\}$, generated as follows:
\begin{itemize}
    \item $X_1$ is uniform$(1,2)$
    \item $X_2$ is uniform$(1,2)$
    \item $X_3$ is ordinal on a grid of $(0,1)$ taking on the $11$ values $\{0.0,0.1,...,1.0\}$
    \item $X_4$ is binary$(0,1)$
    \item $X_5$ is binary$(0,1)$
\end{itemize}

Since the response $(L,R]$ is generated independently from the
covariates $X_1-X_5$, there does not exist any true association
between the survival outcome and covariates, and the tree algorithms
should not split on any of the covariates; unbiasedness would imply
that if the tree is forced to split it would split with equal
probabilities for all five. There are 10,000 simulation trials in
each setting with sample size N = 200. Table 1 gives the raw counts
of how often each variable was selected as the root split variable
for each setting, along with the $p$-value from a Pearson
Chi-squared test of equality of the chances of splitting on each of
the five covariates. Table \ref{bias.table} shows that the
proposed IC tree exhibits little or no bias. Although the pattern for the
Lognormal distribution is marginally significantly different from
uniformity under heavy censoring, there does not appear to be any
systematic preference for either continuous or binary covariates as
the splitting variable.

\begin{table}[th]
\center
\caption{ Proportion of the time IC trees split on each variable}
\label{bias.table}
\begin{threeparttable}
\scalebox{0.85}{%
\begin{tabular}{ lcccccccccccccc}
\hline
&& \multicolumn{6}{c}{Fixed} &&  \multicolumn{6}{c}{Random }  \\  \cline{3-8}  \cline{10-15}
Distribution && $X_1$ & $X_2$ & $X_3$ & $X_4$ & $X_5$ & $p$-value && $X_1$ & $X_2$ & $X_3$ & $X_4$ & $X_5$ & $p$-value \\  \hline
\textit{Light censoring}&&  &  & & & &  && & &  & &  & \\
Exponential && $2057$ & $1968$ & $2048$ & $1951$ & $1976$ & $0.311$ && $2000$ & $1975$ & $1987$ & $2022$ & $2016$ & $0.943$\\
Lognormal    && $1963$ & $1944$ & $2067$ & $1991$ & $2035$ & $0.272$ && $2024$ & $1940$ & $2055$ & $2047$ & $1934$ & $0.142$\\
Weibull  && $2035$ & $1988$ & $2008$ & $1960$ & $2009$ & $0.817$ && $1988$ & $1990$ & $2031$ & $1990$ & $2001$ & $0.957$\\  \hline
\textit{Heavy censoring}&&  &  & & & &  && & &  & &  & \\
Exponential && $2015$ & $1979$ & $2004$ & $2070$ & $1932$ & $0.277$ && $2043$ & $1965$ & $2018$ & $1960$ & $2014$ & $0.627$\\
Lognormal && $2037$ & $2024$ & $1916$ & $2077$ & $1946$ & $0.063$ && $2073$ & $1955$ & $2007$ & $2049$ & $1916$ & $0.077$\\
Weibull && $1975$ & $2099$ & $1975$ & $1946$ & $2005$ & $0.136$ && $1983$ & $2024$ & $2037$ & $1946$ & $2010$ & $0.622$\\ \hline
\end{tabular}
}
 \begin{tablenotes}[para,flushleft]
\footnotesize{Reported values are the number of times the covariate was the root split in 10,000 simulation trials, $p$-value\\ refers to the Chi-squared test of equality of the chances of splitting on each of the five covariates.}
  \end{tablenotes}
\end{threeparttable}
\end{table}

\subsection{Recovering the correct tree structure \label{recoveringtree}}

We next explore the proposed tree's ability to recover the correct underlying tree structure of the data. The simulation setup is as follows.

\begin{figure}[t]
  \begin{center}
   \captionsetup{justification=centering}
      \includegraphics[width=4.5 in,height =3 in]{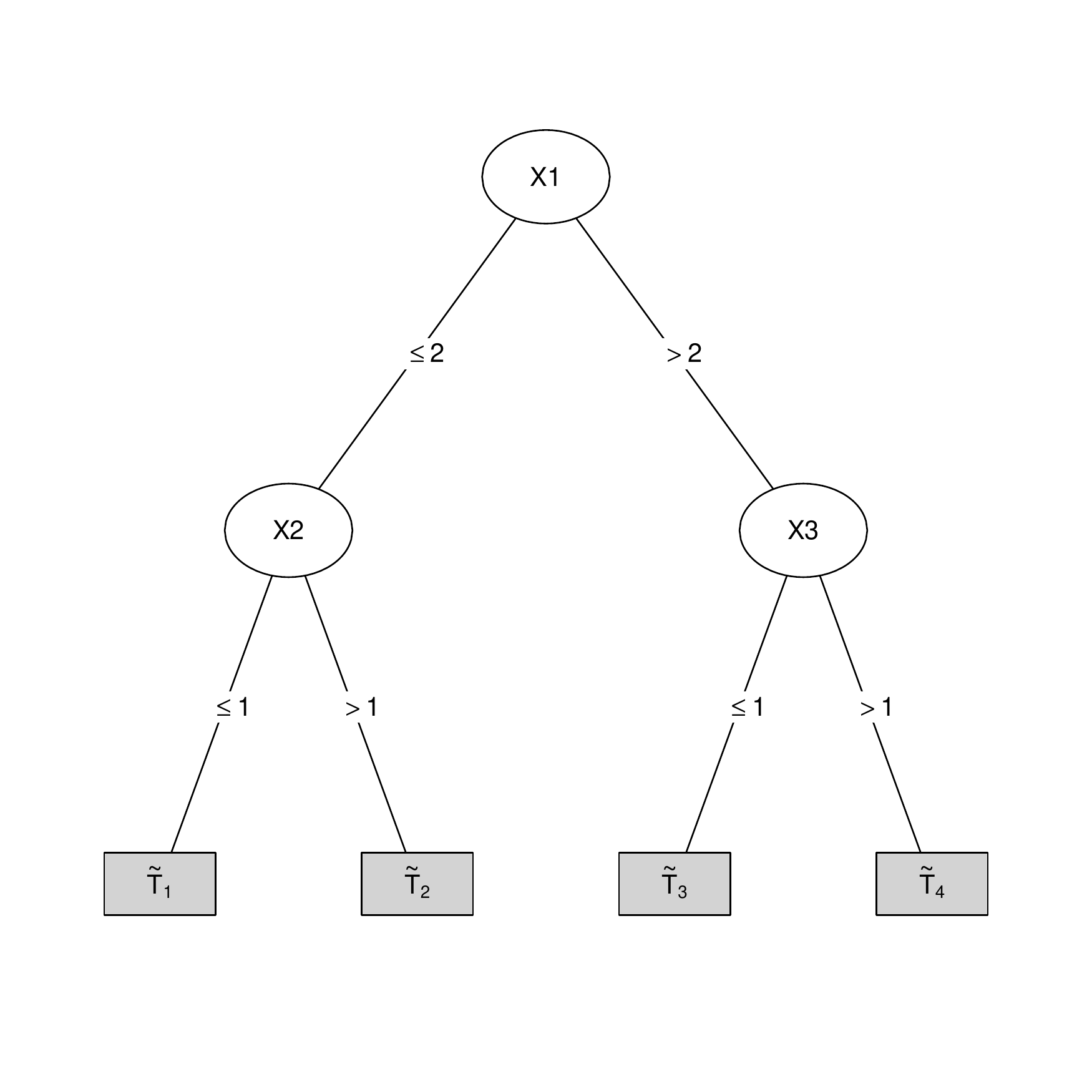}
   \end{center}
     \caption{\label{tree-struct} Tree structure used in simulations of Section \ref{recoveringtree}}
 \end{figure}

There are six covariates $X_1,...,X_6$, where $X_1, X_4$ randomly take values from the set $\{1, 2, 3, 4, 5\}$, $X_2, X_5$ are binary$\{1, 2\}$ and $X_3,X_6$ are $U[0, 2]$. Only the first three covariates $X_1, X_2, X_3$ determine the distribution of the survival (event) time $T$. The survival time $T$ has distribution according to the values of $X_1, X_2, X_3$ by the structure given in Figure \ref{tree-struct}.

We generate $T$ from $5$ different distributions:
\begin{itemize}
    \item Exponential with four different values of $\lambda$ from $\{0.1, 0.23, 0.4, 0.9\}$.
    \item Weibull distribution with shape parameter $\alpha = 0.9$, which corresponds to decreasing hazard with time. The scale parameter $\beta$ takes the values $\{7.0, 3.0, 2.5, 1.0\}$.
    \item Weibull distribution with shape parameter $\alpha = 3$, which corresponds to increasing hazard with time. The scale parameter $\beta$ takes the values $\{2.0, 4.3, 6.2, 10.0\}$.
    \item Log-normal distribution with location parameter $\mu$ and scale parameter $\sigma$ with $4$ different pairs $(\mu, \sigma) = \{(2.0, 0.3), (1.7, 0.2), (1.3, 0.3), (0.5, 0.5)\}$.
    \item Bathtub-shaped hazard model \citep{hjorth1980reliability}. The survival function is given by
    \[ S(t;a,b,c) = \frac{\exp(-\frac{1}{2}at^2)}{(1+ct)^{b/c}} \]
    with $b=1$, $c=5$ and $a$ set to take value $\{0.01, 0.15, 0.20, 0.90\}$.
\end{itemize}

 \noindent

We use the censoring mechanism from the previous section to generate the censoring interval for each
generated $T$, with $\delta_t \in U[0.3,0.7]$. To see the impact of right-censoring, we also consider different percentages of right-censoring among the training data. Specifically, we simulate data without right-censoring ($0\%$ observations right-censored), light censored data with about $20\%$ observations right-censored, and heavy censored data with about $40\%$ observations right-censored. We vary the number $k$ to achieve the desired right-censoring proportion.

We also fit the survival trees of \cite{Hothorn06} with imputed survival times at the beginning, middle and end of the censoring interval for interval-censored observations. The ``oracle'' survival tree of \cite{Hothorn06}, which is fitted using the true event time $T$ (without interval-censoring), is also included, as that represents a reasonable target for the trees addressing interval censoring in each setting.

We run 1,000 simulation trials for each setting to see how well the proposed IC tree recovers the correct tree structure. Table \ref{struct_recover} gives the percentage of the time the correct tree structure is found for each setting.

From Table \ref{struct_recover}, we can see that the common \textit{ad hoc} approach in the literature of imputing the event time at the beginning, middle or end of the censoring interval does not greatly affect the performance of a tree in terms of recovering the correct data structure. In fact, there is virtually no difference between the proposed IC tree and the survival trees fitted with imputed event times. This result holds regardless of the percentage of right-censoring.

There are several possible explanations for this result. One explanation is that the nonparametric maximum likelihood estimator of the survival curve of interval-censored data is only unique up to the so-called equivalence set $(q_i,p_i], i=1,...,n$. In a simple setting where each censoring interval $(L_i,R_i], i=1,...,n$ is non-overlapping, the equivalence set $(q_i,p_i]=(L_i,R_i]$ $\forall i$ where $L_j \leq R_j < L_{j+1} \leq \cdots$. The maximum likelihood estimator demands that the curve be flat between $R_j$ and $L_{j+1}$, and can only jump within the equivalence sets. However, any curve that jumps an appropriate amount within the equivalence class will yield the same likelihood \citep{lindsey1998methods}. In our case, the imputation at the beginning, middle and end of the censoring interval $(L_j, R_j]$ means the corresponding curve jumps at $L_j$, $(L_j+R_j)/2$ and $R_j$, respectively, and those curves are equivalent from the interval-censoring point of view, since they are all the maximum likelihood estimator. It is therefore not surprising that the resulting trees have similar forms since all of the imputation schemes result in curves that provide the same information for the tree to distinguish different survival distributions.

Another explanation is that although imputation at the beginning or end of the interval may bias the estimated survival curves on the terminal nodes of Figure \ref{tree-struct}, the bias amount may be similar for each terminal node. Therefore, the biased curves may be as easily separable as the unbiased curves, which results in similar performance in terms of recovering the correct tree structure. However, such bias may result in worse prediction performance for the trees with imputation, as we will see in the next section.

The proposed IC tree, along with the imputed survival trees, has good performance in recovering the correct tree structure when the sample size is reasonably large. In fact, in the case without right-censoring, the proposed IC tree and the imputed survival trees perform as well as the optimal tree. This demonstrates that interval-censoring has minimal impact on the tree's ability to recover the correct data structure. In contrast, right-censoring has significant effect on the tree's ability to recover the correct data structure, as the performance deteriorates when the right-censoring proportion increases.
%------------------------------------------------------------
\begin{table}[t]
\center
\caption{\label{struct_recover} Tree structure recovery rate in percentage.}
\scalebox{0.85}{%
\hspace*{-1cm}
\begin{threeparttable}
\begin{tabular}{lrrrrrrrrrrrrrrrrrrr}
  \hline
 \multicolumn{2}{c}{N=200} && \multicolumn{5}{c}{\textit{$0\%$ right-censoring}} & \multicolumn{5}{c}{\textit{Light censoring}} && \multicolumn{5}{c}{\textit{Heavy censoring}}  \\  \cline{3-7}  \cline{9-13}  \cline{15-19}
 Distribution && oracle & IC & L & M & R && oracle & IC & L & M & R && oracle & IC & L & M & R \\
  \hline
Exponential && 53.2 & 53.4 & 51.6 & 52.0 & 52.5 && 51.3 & 37.0 & 36.2 & 35.5 & 35.5 && 52.3 & 16.7 & 15.8 & 15.9 & 16.3 \\
  Weibull\_I && 87.8 & 87.4 & 86.7 & 86.5 & 86.9 && 84.7 & 80.1 & 80.2 & 80.2 & 78.9 && 83.8 & 37.5 & 36.6 & 36.7 & 34.0 \\
  Weibull\_D && 46.6 & 46.3 & 47.1 & 46.0 & 46.8 && 43.9 & 28.8 & 28.4 & 29.0 & 28.4 && 43.4 & 16.3 & 15.7 & 16.9 & 16.6 \\
  Lognormal && 86.3 & 85.6 & 86.0 & 86.4 & 86.8 && 85.5 & 76.3 & 76.5 & 76.9 & 75.8 && 87.4 & 12.3 & 12.1 & 11.3 & 9.4 \\
  Bathtub && 43.4 & 53.7 & 50.4 & 44.2 & 44.4 && 40.6 & 22.1 & 20.0 & 16.7 & 15.8 && 42.2 & 6.5 & 7.3 & 6.0 & 5.2 \\
   \hline
\end{tabular}
 \begin{tablenotes}[para,flushleft]
\footnotesize{Numbers in the table show the percentage of the time the correct tree structure is recovered in 1,000 simulation trials. ``Oracle'' denotes conditional inference survival tree results using the actual survival time $T$ and ``IC'' denotes the proposed tree result, while ``L'', ``M'', and ``R'' denote the left, middle and right imputation-based tree result, respectively}
  \end{tablenotes}
\end{threeparttable}
}
\end{table}

A surprising anomaly is that the IC tree outperforms the oracle tree for the bathtub distribution when there is no right-censoring. Almost always this corresponds to the oracle tree not making one of the two splits on the second level of the tree. The bathtub survival functions corresponding to these splits are close for small failure times, and apparently in that situation the \cite{turnbull1976empirical} estimates in that region are sometimes far enough apart for the splitting test to reach significance at a .05 level when the KM-based test does not. This behavior disappears when the cutoff of the test is set to $\alpha=.10$ rather than .05, or if the sample size is increased to roughly 300.

\subsection{Prediction performance}
We use three simulation setups to test the prediction performance of the proposed IC tree. To see how it compares with a (semi-)parametric model, we also include the Cox proportional hazards model implemented in the \texttt{R} package \texttt{icenReg} \citep{icenReg} in the simulations for comparison. To see the amount of information loss due to interval-censoring, we include the oracle versions of both tree and Cox models, which are fitted using the actual event time $T$. Also included are the survival trees and Cox PH models with imputation at the beginning, middle and end of the censoring interval for interval-censored observations. The three survival families are as follows:

\begin{enumerate}[label=(\roman*)]
    \item Tree structured data as in Section \ref{recoveringtree};
    \item  $\vartheta = -X_1-X_2$;
    \item $\vartheta = -\left[cos\left(\left(X_1+X_2\right)\cdot \pi\right)+\sqrt{X_1+X_2}\right]$,
\end{enumerate}
where $\vartheta$ is a location parameter whose value is determined by covariates $X_1$ and $X_2$. In the first setup, data are generated according to the tree structure described in Section \ref{recoveringtree}, so the trees should perform well. In this setup the five survival distributions used in Table \ref{struct_recover} are again used. The second and third setups are similar to those in \cite{hothorn2004bagging}. In these settings six independent covariates $X_1,...,X_6$ serve as predictor variables, with $X_2, X_3,X_6$ binary$\{0,1\}$ and $X_1, X_4, X_5$ uniform$[0,1]$. The survival time $T_i$ depends on $\vartheta$ with three different distributions:

\begin{itemize}
    \item Exponential with parameter $\lambda = e^{\vartheta}$;
    \item Weibull with increasing hazard, scale parameter $\lambda = 10e^{\vartheta}$ and shape parameter $k=2$;
    \item Weibull with decreasing hazard, scale parameter $\lambda = 5e^{\vartheta}$ and shape parameter $k=0.5$.
\end{itemize}

In the second setup where $\vartheta = -X_1-X_2$, the linear proportional hazards assumption is satisfied, so the Cox PH model should perform best in this setup. The third setup is similar to the second except that $\vartheta$ in this setup has a more complex nonlinear structure in terms of covariates, which makes the distributions of $T_i$ satisfy neither the Cox PH model nor the tree structure. Such a setup is to test how effective the IC trees and Cox PH model are in a real world application where survival time might have a complex structure.

In all setups, we use the censoring mechanism described earlier to generate the censoring interval with $\delta_t$ randomly generated from the uniform distribution $U[0.3,0.7]$. Corresponding results using a wider censoring interval $U[1.0,1.3]$ gave similar results, except that all methods other than the oracle methods had higher predictive error because of the greater uncertainty caused by the wider censoring interval. Three possible right-censoring rates, $0\%$ right-censoring, light censoring with about $20\%$ observations being right-censored and heavy censoring with about $40\%$ observations being right-censored, are considered in each setting.

The survival time $T_i$ in the test set is also generated according to this process, except that no censoring is used, i.e. the survival time $T_i$ is never censored. The test set is set to have the same sample size as the training set. The size $N = 200$ is used in the simulations presented here; results with $N=400$ were similar.

To compare different methods, we use the average integrated $L_2$ difference between the true and estimated survival curves for observations in the test set,
\[ \frac{1}{N}\sum_{i=1}^N \frac{1}{{\rm max}_j{(T_j)}} \int^{{\rm max}_j{(T_j)}}_0 [\hat S_i(t) - S_i(t)]^2 dt,\]
where $T_j$ is the (actual) event time of the $j$th observation in the test set and $\hat S_i(\cdot)$ ($S_i(\cdot)$) is the estimated (true) survival function for the $i$th observation. The most popular measure of error in the survival context is the (integrated) Brier score introduced by \cite{graf1999assessment}, and comparing methods using $L_2$ difference is equivalent to using the expected value of the Brier score. The key is to estimate the survival function $\hat{S}(t)$, which is estimated by the NPMLE curves in each node for the trees and $\hat{S}(t) = \hat{S}_0(t)^{e^{X \hat{\beta}}}$ for the Cox model. As long as $\hat{S}(t)$ is produced, we can use it to compute the integrated $L_2$ difference.

\begin{sidewaysfigure}
\includegraphics[width=\textwidth]{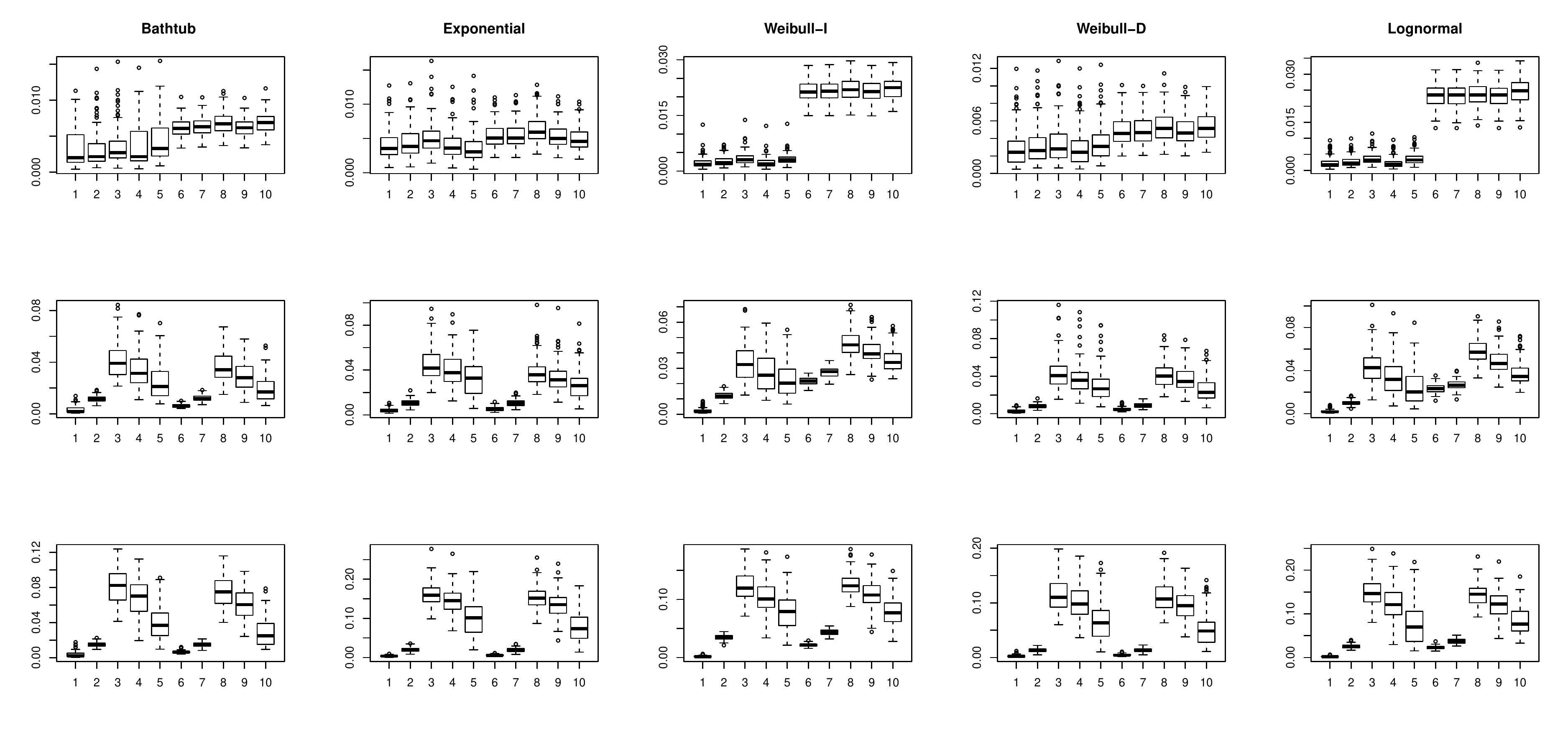}
\caption{\label{fig:Boxplot1}Setting $1$: integrated $L_2$ difference boxplots with $N=200$. Methods are numbered as
$1$-Oracle survival tree, $2$-IC tree, $3$-Survival tree with left imputation, $4$-Survival tree with midpoint imputation, $5$-Survival tree with right imputation, $6$-Oracle Cox model, $7$-Cox model for interval-censored data, $8$-Cox model with left imputed data, $9$-Cox model with midpoint imputed data, $10$-Cox model with right imputed data. Top row are results without right-censoring, middle row are results of light (right) censoring and bottom row are results of heavy (right) censoring.}
\end{sidewaysfigure}

\begin{sidewaysfigure}
\includegraphics[width=\textwidth]{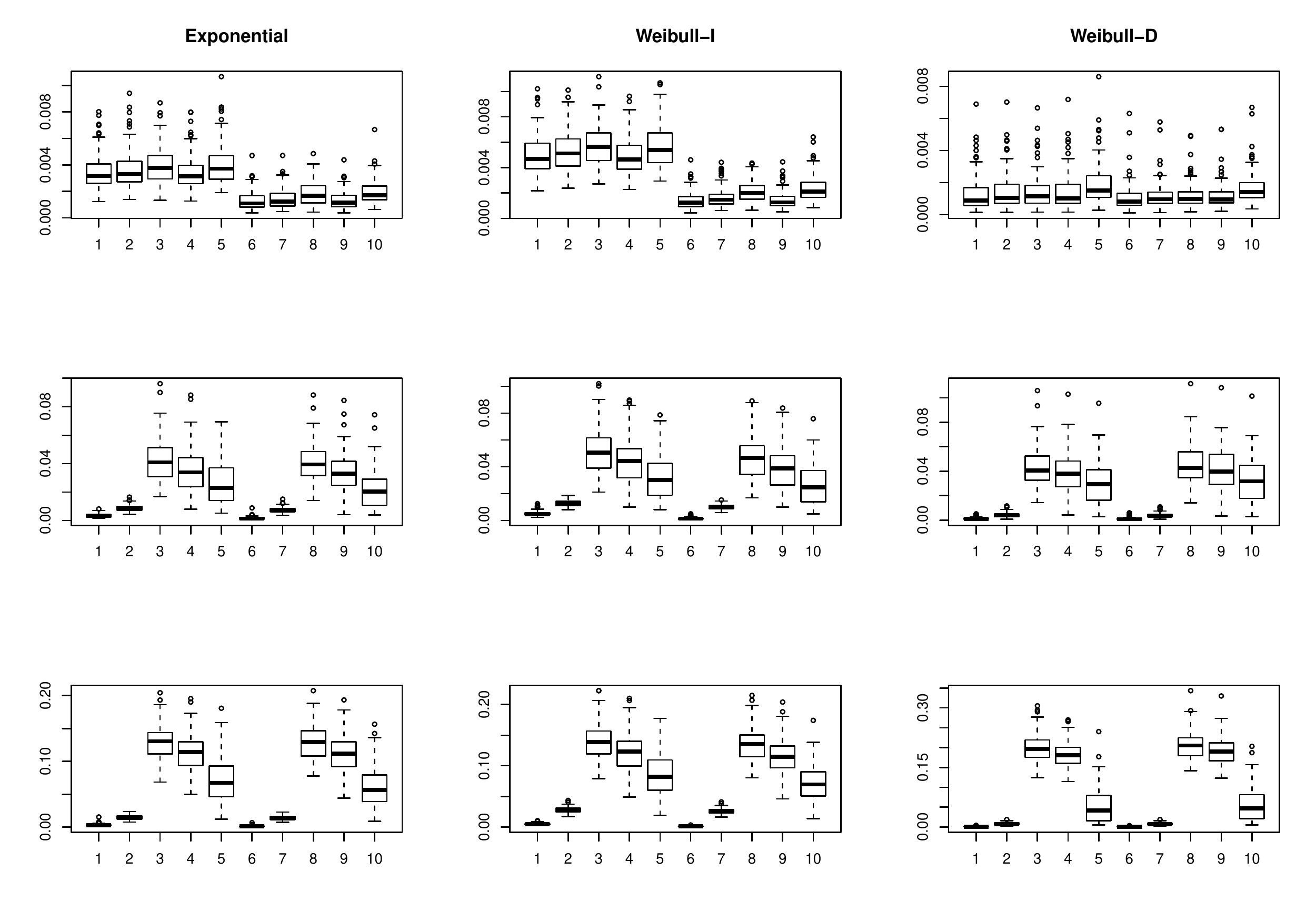}
\caption{\label{fig:Boxplot2}Setting $2$: integrated $L_2$ difference boxplots with $N=200$. Methods are numbered as
$1$-Oracle survival tree, $2$-IC tree, $3$-Survival tree with left imputation, $4$-Survival tree with midpoint imputation, $5$-Survival tree with right imputation, $6$-Oracle Cox model, $7$-Cox model for interval-censored data, $8$-Cox model with left imputed data, $9$-Cox model with midpoint imputed data, $10$-Cox model with right imputed data. Top row are results without right-censoring, middle row are results of light (right) censoring and bottom row are results of heavy (right) censoring.}
\end{sidewaysfigure}

\begin{sidewaysfigure}
\includegraphics[width=\textwidth]{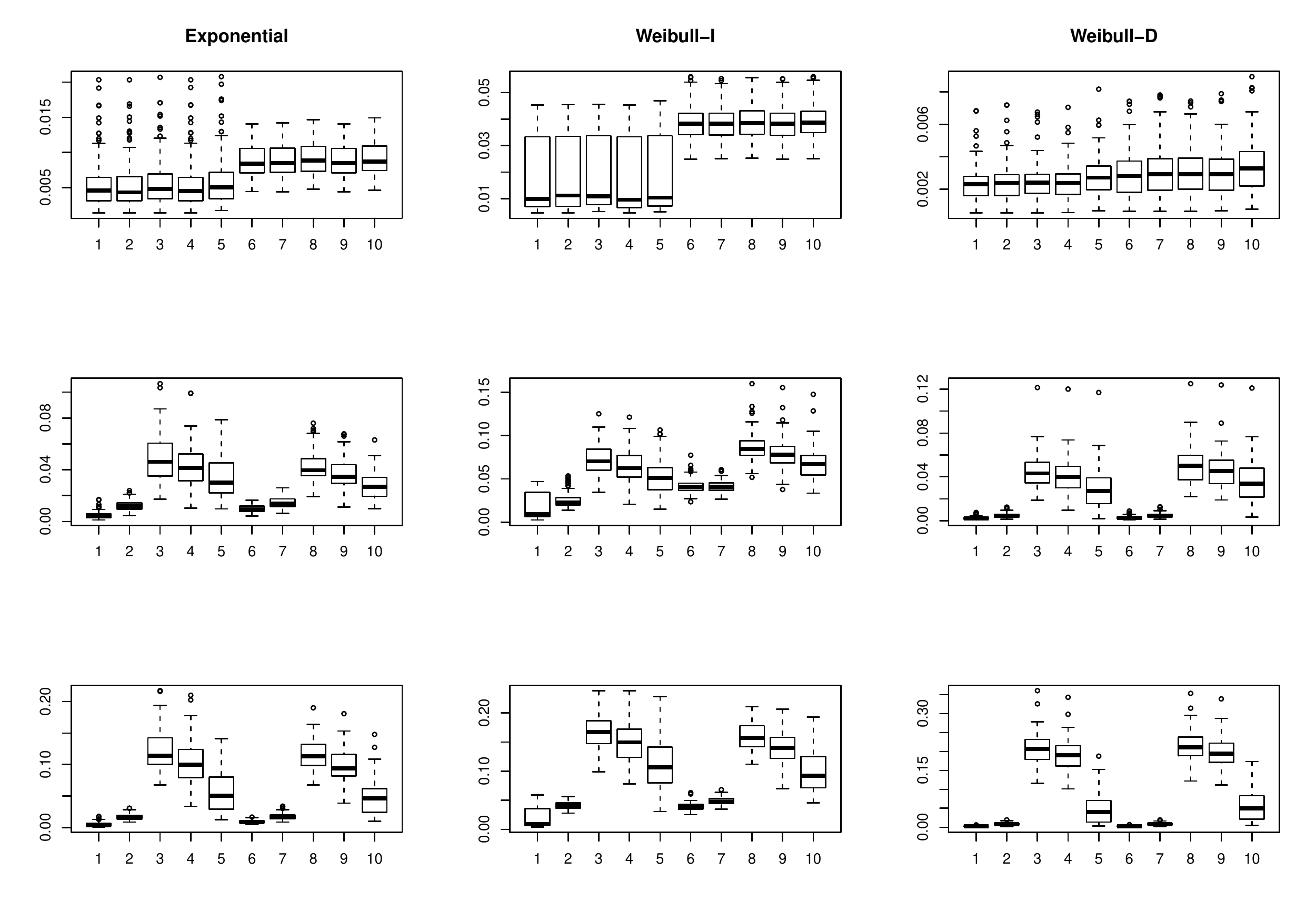}
\caption{\label{fig:Boxplot3}Setting $3$: integrated $L_2$ difference boxplots with $N=200$. Methods are numbered as
$1$-Oracle survival tree, $2$-IC tree, $3$-Survival tree with left imputation, $4$-Survival tree with midpoint imputation, $5$-Survival tree with right imputation, $6$-Oracle Cox model, $7$-Cox model for interval-censored data, $8$-Cox model with left imputed data, $9$-Cox model with midpoint imputed data, $10$-Cox model with right imputed data. Top row are results without right-censoring, middle row are results of light (right) censoring and bottom row are results of heavy (right) censoring.}
\end{sidewaysfigure}

Figures \ref{fig:Boxplot1}--\ref{fig:Boxplot3} give side-by-side integrated $L_2$ difference
boxplots for all three setups with sample size $N=200$. Signed-rank
tests show that any differences in the figures are statistically
significant. Figure \ref{fig:Boxplot1} shows that in the presence of
right-censoring the proposed IC tree has the best prediction
performance (except for the oracle methods) in the first setup where
the true structure is a tree. The proposed IC tree also outperforms
the IC Cox model in the third setup (Figure \ref{fig:Boxplot3}),
highlighting the ability of the tree to mimic a complex structure
because of its flexible nature. The biggest advantage of the IC tree
over the IC Cox model occurs for the Weibull survival distribution
with increasing hazard and the lognormal distribution. As expected,
the IC Cox model usually outperforms the IC tree in the second setup
(Figure \ref{fig:Boxplot2}), but performance is actually comparable
from a practical point of view (and the trees can be better than the
Cox models for a Weibull survival distribution with increasing
hazard), illustrating that the tree can even represent a linear
model reasonably.

Although the imputation-based methods seem to have comparable ability to recover the correct data structure as does the IC tree as we have seen in Section \ref{recoveringtree}, these methods are noticeably worse in terms of prediction in the settings with right-censoring. We can see that the proposed IC tree has smaller $L_2$ difference than the trees with imputed data in all such settings. In contrast, the IC tree has no significant difference from the imputed survival trees when there is no right-censoring (indeed, all of the methods have comparable performance). This pattern is driven by the poor performance of the Kaplan-Meier curves used at the terminal nodes of the imputation-based trees to estimate the upper tail of the survival distribution compared to the \cite{turnbull1976empirical} estimator used in the IC tree. This difference disappears when there is no right-censoring, resulting in nearly identical performance for all methods. An interesting observation is that right endpoint imputation results in better prediction performance than imputation at the beginning or middle points of the censoring interval, since this pushes uncensored observations further into the tail, even though endpoint imputation results in the worst performance in terms of recovering the correct tree structure (as seen in section $3.2$). End-point imputation also works best in terms of prediction for the Cox model.

The relative performance of the IC tree to the oracle tree is similar to the relative performance of the IC Cox model to the oracle Cox model. This suggests that information loss due to interval-censoring has a similar effect on the tree and Cox models in terms of prediction performance. This also means that the relative performance of the IC tree and the IC Cox model depends on the relative performance of their corresponding oracle versions, and the general conclusions regarding the performance of survival trees and the Cox model carry over to the interval-censoring situation.

For both the IC tree and the Cox model for interval-censored data, performance is relatively unchanged when the right-censoring proportion increases. However, the imputation-based trees and Cox models are quite sensitive to right-censoring as we can see their performances deteriorate dramatically when the right-censoring percentage increases.

The iterative nature of the NPMLE of \cite{turnbull1976empirical} makes it much more computationally intensive than is the Kaplan-Meier estimator, and as a result the IC tree takes considerably longer to calculate than does an ordinary conditional inference survival tree. On a computer running the 64 bit Windows 7 Professional operating system with a 3.40 GHz processor and 8.0GB of RAM the calculation of the IC tree in a single run in the simulations of Section \ref{recoveringtree} for light right-censoring takes roughly 0.1 to 0.3 seconds for $N=50$, 0.5 to 1.1 seconds for $N=100$, 2 to 4 seconds for $N=200$, and 10 to 16 seconds for $N=400$ (suggesting a multiplicative relationship where doubling sample size implies roughly quadrupling computation time), with this being driven almost completely by calculation of the NPMLE (in contrast, the imputation-based tree averages less than 0.02 seconds in computation time in all cases). Use of \texttt{Microsoft R Open} (\texttt{https://mran.microsoft.com/open/}), with its much faster math libraries, can cut computation time of the IC tree (often being 10 to 30\% faster in the situations examined here).

\section{Real data example}
The Signal Tandmobiel$^{\mbox{\textregistered}}$ study is a longitudinal
prospective oral health study that was conducted in the Flanders region of Belgium from
1996 to 2001. In this study, $4430$ first year primary school
schoolchildren were randomly sampled
at the beginning of the study and were dental-examined annually by
trained dentists. The data consist of at most six dental
observations for each child including time of tooth emergence,
caries experience and data on dietary and oral hygiene habits. The
details of study design and research methodology can be found in
\cite{vanobbergen2000signal}. The data are provided as the {\tt
tandmob2} data set in the {\tt R} package {\tt bayesSurv}
\citep{bayesSurv}. The response variable examined is the time to emergence of the  permanent upper left first premolar (tooth $24$ in European dental notation; this emergence time was also investigated in \citeauthor{gomez2009tutorial}, \citeyear{gomez2009tutorial}). Since permanent teeth do not emerge before age $5$, the origin (zero) time is set at $5$ years of age throughout, as suggested in \cite{intervaloverview}. Potential predictors of emergence time of the child's tooth include gender, province, evidence of fluoride intake, type of educational system, starting age of brushing teeth, the total number of deciduous teeth that were decayed or missing due to caries or filled ({\tt DMF.Score}), and the total number of deciduous teeth that were removed because of orthodontic reasons ({\tt BAD.Score}).

\begin{sidewaysfigure}
\includegraphics[width=\textwidth]{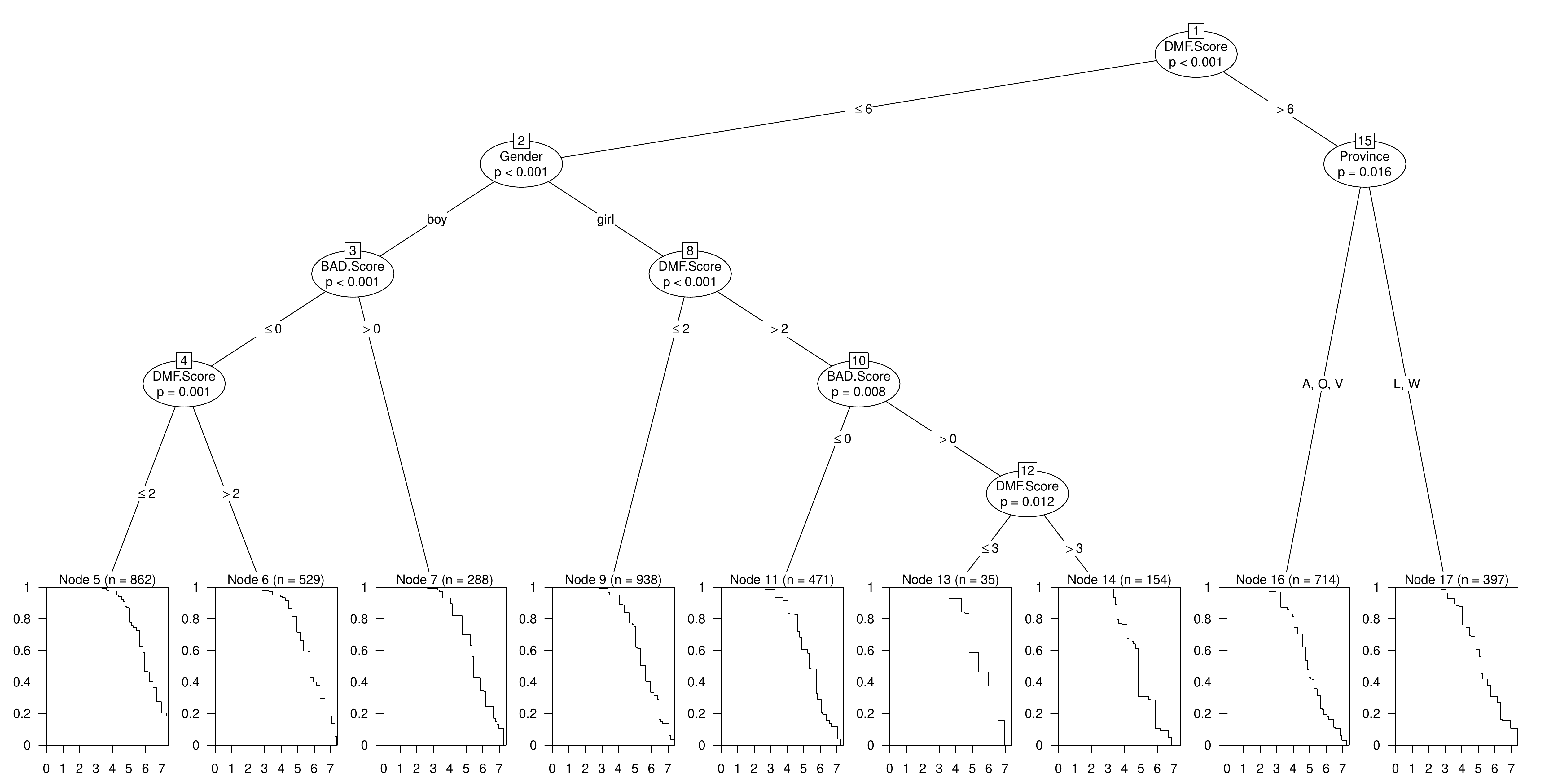}
\caption{\label{example24}Fitted IC tree for the emergence time in years after
the age of 5 of the permanent upper left first premolar tooth.}
\end{sidewaysfigure}

Figure \ref{example24} gives the IC tree for the emergence time of the tooth. In addition to the interval-censoring, 37\% of the observations for this variable are right-censored. The tree is laid out with emergence time distributions corresponding to later emergence on the left to earlier emergence on the right. We can see that more general decay is strongly associated with earlier emergence time, and girls tend to have earlier emergence times; these patterns were also noted by \cite{emergence} and \cite{intervaloverview}. We can also note that more orthodontic removal of deciduous teeth is associated with earlier emergence time; this could reflect a reverse causality in that children are more likely to have deciduous teeth removed to make room for permanent teeth in a not-fully-developed jaw when a dental professional sees that a permanent tooth is about to emerge earlier than is typical (personal communication, Keith B. Annapolen, D.D.S., February 20, 2017). One other interesting pattern is that for children with a large amount of decay ($\mbox{DMF.Score}>6$) their province of residence is predictive; comparison of the estimated survival distributions indicates earlier emergence times in Antwerp (Antwerpen, A), East Flanders (Oost Vlaanderen, O), and Flemish Brabant (Vlaams Brabant, V), and later emergence times in Limburg (L) and West Flanders (West Vlaanderen, W). This location-based pattern has been noted before in these data, and apparently is due to the combination of misclassification of caries experience by certain examiners and the fact that examiners worked in geographic areas close their homes. See \cite{ordinallogistic} and \cite{flexibleAFT} for further discussion.

A major benefit of constructing a tree is that it highlights interaction effects in an obvious way. According to the tree when a child has less decay experience ($\mbox{DMF.Score}\le 6$) several covariates are predictive for emergence time (gender, decay, and orthodontic tooth removal), but for heavy decay experience ($\mbox{DMF.Score}>6$) only province is predictive. The existence of this pattern (which is of course not uncovered using standard analyses without prior knowledge to look for it) is confirmed by Cox regression fits to these two subgroups, as given in Table \ref{subgroupfits}. The observed test statistics are based on bootstrap estimates of the standard errors of the coefficients, and effect codings are used to code the two multiple-category predictors  Province and Education. It is evident that the proportional hazards fits are consistent with the tree fit, as gender, DMF.Score, and BAD.Score are the only variables that are statistically significant for the lower decay group, and several province effect codings are the only variables that are statistically significant in the higher decay group.

\begin{table}[th]
\center
\caption{\label{subgroupfits} Cox model fits for emergence time of the permanent upper left first premolar, separated by $\mbox{DMF.Score}\le > 6$.}
\begin{tabular}{lcccc}
\multicolumn{5}{c}{$\mbox{DMF.Score}\le 6$} \\
\multicolumn{1}{c}{\it Predictor} & {\it Coefficient} & {\it s.e.(Coeff.)} & $z$ & $p$  \\
Gender (girl) &   0.45850   &  0.04930 & 9.3010 & $< .001$ \\
Antwerpen &    -0.01900 &     0.04271 &   -0.4448 &  .657 \\
Limburg  &    -0.05273 &     0.05338 &  -0.9879 &  .323 \\
Oost Vlaanderen &     -0.04915 &    0.05737 &  -0.8567 &  .392 \\
Vlaams Brabant &      0.09356 &     0.06397 &   1.4630 &  .144 \\
West Vlaanderen &     0.02731 &    0.05417 &   0.5042 &  .614 \\
Fluorosis  &        -0.04817   &    0.06779 &  -0.7106 &   .477 \\
Community educ. &   0.04242   &     0.05881 &   0.7213 &  .471 \\
Free educ. &   -0.05450   &    0.03668 &  -1.4860 &  .137 \\
Province educ. &    0.01208  &     0.05350 &   0.2258 &  .821 \\
Started brushing  &       0.03416 &     0.02581  &  1.3230 &  .186 \\
DMF.Score     &  0.05434  &   0.01314  &  4.1360 &  $< .001$ \\
BAD.Score &      0.10630 &    0.03394 &   3.1340 &  $< .001$  \\[3ex]
\multicolumn{5}{c}{$\mbox{DMF.Score}> 6$} \\
\multicolumn{1}{c}{\it Predictor} & {\it Coefficient} & {\it s.e.(Coeff.)} & $z$ & $p$  \\
Gender (girl)  &  0.04096  &    0.07828 &   0.5232  & 0.601 \\
Antwerpen  &   0.11470  &    0.06944  &  1.6510  & 0.099 \\
Limburg   &  -0.12580  &    0.09071  & -1.3860  & 0.166 \\
Oost Vlaanderen  &   0.16310  &    0.06898  &  2.3640  & 0.018 \\
Vlaams Brabant  &   0.03150  &  0.10080   & 0.3124  & 0.755 \\
West Vlaanderen  &  -0.18350  &   0.09447  & -1.9420  & 0.052 \\
Fluorosis   &       0.09043  &    0.13580  &  0.6660  & 0.505 \\
Community educ. &    0.04565  &    0.08979  &  0.5084  & 0.611 \\
Free educ.  &   0.03078   &   0.05728  &  0.5374  & 0.591 \\
Province educ.  &  -0.07643   &   0.08297  & -0.9212  & 0.357 \\
Started brushing   &    0.04507   &   0.03349   & 1.3460  & 0.178 \\
DMF.Score  &   0.02311  &    0.03421  &  0.6755  & 0.499 \\
BAD.Score  &   0.04017   &   0.02443  &  1.6440  & 0.100
   \end{tabular}
\end{table}

\section{Conclusion}

In this paper, we have proposed a new tree algorithm that is
designed to handle interval-censored data. Through simulation study,
we see that the proposed IC tree inherits the unbiasedness property
from the conditional inference tree of \cite{Hothorn06}, and
performs well in terms of both recovering the correct data structure
and prediction performance. In the presence of right-censoring the
IC tree is more effective than the IC Cox proportional hazards model
fit when the true model is not a linear model and comparable to it
when it is, and outperforms survival trees based on imputed data. An
interesting question for future work is whether this method can be
adapted to doubly-interval-censored data, in which both the start time
and the end time of a time to event are interval-censored.

The discussion here is based on the assumption that the process generating the censoring intervals is independent of both the covariates and the survival times. A referee has pointed out that the proposed tree has a potential advantage over an imputation-based tree in the less strict situation where the interval-generating process is independent of the survival times given the covariates. \cite{FayShih} showed that in this circumstance permutation tests based on right endpoint imputation have greater than nominal Type I error rates, while tests based on interval-censored log-rank scores have close to nominal coverage. Given that the splitting rule of conditional inference trees is based on the $p$-value of a permutation test, this suggests that the similar tree recovery performance of the different types of trees noted here in Section \ref{recoveringtree} might not carry over to this less-restrictive situation, with the interval-censored tree exhibiting better behavior.

The proposed method is implemented in the \texttt{R} package \texttt{LTRCtrees} \citep{LTRCtrees}.

\section*{Acknowledgments}

We would like to thank two referees for helpful comments that improved the quality of the paper. Data collection of the Signal Tandmobiel$^{\mbox{\textregistered}}$
data was supported by Unilever, Belgium. The Signal-Tandmobiel
project comprises the following partners: Dominique Declerck
(Department of Oral Health Sciences, KU Leuven), Luc Martens (Dental
School, Gent Universiteit), Jackie Vanobbergen (Oral Health
Promotion and Prevention, Flemish Dental Association \& Dental
School, Gent Universiteit, Peter Bottenberg (Dental School, Vrije
Universiteit Brussel), Emmanuel Lesaffre (L-Biostat, KU Leuven), and
Karel Hoppenbrouwers (Youth Health Department, KU Leuven; Flemish
Association for Youth Health Care). We thank Keith Annapolen,
Arno\u{s}t Kom{\'a}rek, and Emmanuel Lesaffre for helpful discussion
of this material.

\bibliographystyle{apalike}

\end{document}